\documentclass[journal=jctcce,manuscript=article]{achemso}
\usepackage[version=3]{mhchem} 
\usepackage{longtable}
\usepackage[table]{xcolor}

\author{Fabio Caruso}
\affiliation{Department of Materials, University of Oxford, Parks Road, Oxford OX1 3PH, United Kingdom}
\author{Matthias Dauth}
\affiliation{Theoretical Physics IV, University of Bayreuth, D-95440 Bayreuth, Germany}
\author{Michiel J. van Setten}
\affiliation{Nanoscopic Physics, Institute of Condensed Matter and Nanosciences, Universit\'{e} Catholique de Louvain, 1348 Louvain-la-Neuve, Belgium}
\author{Patrick Rinke}
\email{patrick.rinke@aalto.fi}
\affiliation{COMP/Department of Applied Physics, Aalto University, P.O. Box 11100, Aalto FI-00076, Finland}
\title[Benchmark of \textit{GW} approaches]{Benchmark of \textit{GW} approaches for the \textit{GW}100 testset}

\begin{document}
\begin{abstract}
For the recent \textit{GW}100 test set of molecular ionization energies, we present a comprehensive assessment of different $GW$ methodologies: fully self-consistent $GW$ (sc$GW$), quasiparticle self-consistent $GW$ (qs$GW$), partially self-consistent $GW_0$ (sc$GW_0$), perturbative $GW$ ($G_0W_0$) and optimized $G_0W_0$ based on the minimization of the deviation from the straight-line error (DSLE-minimized $GW$). We compare our $GW$ calculations to coupled-cluster singles, doubles, and perturbative triples [CCSD(T)] reference data for \textit{GW}100. 
We find sc$GW$ and qs$GW$ ionization energies in excellent agreement with CCSD(T), with discrepancies typically smaller than 0.3~eV (sc$GW$) respectively 0.2~eV (qs$GW$). For sc$GW_0$ and $G_0W_0$ the deviation from CCSD(T) is strongly dependent on the starting point. We further relate the discrepancy between the $GW$ ionization energies and CCSD(T) to the deviation from straight line error (DSLE). 
In DSLE-minimized $GW$ calculations, the DSLE is significantly reduced, yielding a systematic improvement in the description of the ionization energies. 
\end{abstract}


\section{Introduction}

Many-body perturbation theory provides an ideal framework 
for the first-principles study of electronic excitations 
in molecules and solids.\cite{fetter}
At variance with approaches based on 
density-functional theory (DFT),\cite{PhysRev.136.B864,PhysRev.140.A1133}
the description of electronic many-body interactions through the electron self-energy 
facilitates a seamless account of exact exchange
and screening, which are essential to predict electronic 
excitations with quantitative accuracy.\cite{Aulbur/Jonsson/Wilkins:2000,Onida2002,Rinke/2005,Faber/etal:2014}
The $GW$ approximation\cite{Hedin1965,Hybertsen1986} 
provides an ideal compromise between accuracy and computational 
cost and it has, thus, evolved into the state-of-the-art technique 
for the computation of ionization energies and band gaps 
in molecules and solids.\cite{Onida2002}

$GW$ calculations are typically based on first-order perturbation theory ($G_0W_0$),\cite{Hybertsen1986} a procedure 
that introduces a spurious dependence of the results on the starting point, that is, 
the initial reference ground state the perturbation is applied to \cite{Rinke/2005,Fuchs2007,marom/2012,Gallandi/etal:2016,Knight/etal:2016}. The starting-point dependence may be reduced by resorting to partial 
self-consistent approaches,\cite{marom/2012, kaplan15gwso} 
such as eigenvalue self-consistent $GW$ or self-consistent $GW_0$ (sc$GW_0$), 
and it is completely eliminated in the
self-consistent $GW$ method (sc$GW$) \cite{caruso/2012,caruso/2013} 
-- in which the Dyson equation is solved fully iteratively -- 
and in quasi-particle self-consistent $GW$ (qs$GW$) \cite{faleev/2004,vanschilfgaarde/2006,kaplan16}. While sc$GW$ implementations are still relatively rare \cite{stan/2006,stan/2009,kutepov,rostgaard/2010,kutepov2,caruso/2012,caruso/2013, Koval/etal:2014,Wang:2015,Chu/etal:2016}, qs$GW$ is now widely used. \cite{Kotani/etal:2007,Kotani/etal:2007_2,Shiskin/etal:2007,Bruneval2014,bech15book,kaplan16} Moreover, with rare exceptions \cite{Koval/etal:2014}, sc$GW$ and qs$GW$ are typically not implemented in the same code and have therefore not been systematically compared.

Given the various flavors of the self-consistent $GW$ methodology, 
benchmark and validation are important instruments to 
(i)   quantify the overall accuracy of $GW$ calculations;  
(ii)  reveal the effects of different forms of self-consistency;
(iii) identify new ways to improve over existing techniques for quasiparticle calculations. 
The $GW$100 set provides an ideal test-case for addressing these challenges.\cite{vansetten/2015}
This benchmark set is specifically designed to target the assessment of ionization energies
and it is composed of 100 molecules of different bonding types, chemical compositions, and ionization energies.

In this manuscript, we present the ionization energies for the molecules of 
$GW$100 test set calculated with $G_0W_0$, sc$GW_0$, sc$GW$, and qs$GW$. We analyse their behaviour in terms of the change in the electron density, the screening properties and the treatment of the kinetic energy.
The accuracy of different $GW$ approaches is established based on the 
comparison with coupled-cluster singles, doubles, and perturbative 
triples\cite{purvis82, raghavachari89, szabo/1989} 
[CCSD(T)] energies obtained for the same geometries and basis sets.\cite{krause/2015} 
Our study reveals that sc$GW$ and qs$GW$ ionization energies differ on average 
by 0.3~eV and 0.15~eV from the CCSD(T) reference data, respectively. 
The discrepancy of $G_0W_0$ and sc$GW_0$ from CCSD(T), on the other 
hand, is contingent on the starting point. For the $GW$100 set, we 
report an average starting-point dependence of 1 and 0.4 eV for $G_0W_0$ and 
sc$GW_0$, respectively. 
Correspondingly, the starting point introduces an additional degree of freedom 
that allows one to improve the agreement with CCSD(T), e.g., by 
imposing the satisfaction of exact physical constraints. 
One of such constraint is the linearity of the total energy at 
fractional particle numbers.\cite{perdew/1982}
The deviation from straight line error (DSLE) has been shown to 
lead to systematic errors in DFT, such as the tendency to overly 
localize or delocalize the electron density.\cite{YangScience,Atalla/etal:2016}
Within the context of $GW$ calculations, the DSLE may be minimized by varying
the starting point. This procedure we 
refer to as the DSLE-minimized $GW$ approach (DSLE-min).\cite{dauth/2015} 
We show here that DSLE-min $GW$
reduces the discrepancy with CCSD(T) for the $GW$100 set as 
compared to sc$GW$ with an average absolute deviation slightly 
larger than that of qs$GW$ (0.26~eV, based on the def2-TZVPP basis set).
Overall, our results provide a comprehensive assessment of the 
starting-point dependence, the accuracy of $G_0W_0$ and 
self-consistent $GW$ methods, and suggest that the DSLE 
minimization may provide a strategy to improve the accuracy 
of the $GW$ method at the cost of $G_0W_0$ calculations.

The manuscript is organized as follows. 
In Sec.~\ref{sec:method} we review the basics of the $GW$ method and 
self-consistency. The ionization energies for 
the $GW$100 test set are reported in Sec.~\ref{sec:IE}, and discussed in 
Sec.~\ref{sec:g0w0vsSC}.
DSLE-min $GW$ results are discussed in Sec.~\ref{sec:DSLE}. 
Conclusions and final remarks are presented in Sec.~\ref{sec:conc}.

\section{Methods}\label{sec:method}

In the following, we give a brief introduction to the $GW$ methodology 
employed throughout the manuscript: sc$GW$, sc$GW_0$, qs$GW$, 
perturbative $G_0W_0$, and DSLE-min $GW$. 

In the sc$GW$ approach, the interacting Green's function $G$ 
is determined through the iterative solution of Dyson's equation
\begin{equation}\label{eq:dyson}
G^{-1}=G_{\rm 0}^{-1}-[\Sigma-v_{0} + \Delta v_{\rm H}] \quad.
\end{equation}
$\Delta v_{\rm H}$ denotes the change of the Hartree potential,
which accounts for the density difference between $G_0$ and $G$, 
and $v_{0}$ is the exchange-correlation potential of the preliminary calculation.
The non-interacting Green's function $G_0$ may be expressed as
\begin{align}
G_0^{\sigma}({\bf r},{\bf r'},\omega)=\sum_n \frac{\psi_{n\sigma}({\bf r})\psi_{n\sigma}^{*}({\bf r'})}
{\omega-(\epsilon_{n\sigma}-\mu)-i\eta \,sgn(\mu-\epsilon_{n\sigma})}
\end{align}
where $\mu$ is the Fermi energy, and $\eta$ a positive infinitesimal.
$\psi_{n\sigma}$ and $\epsilon_{n\sigma}$ denote a set of single-particle 
orbitals and eigenvalues determined from an independent-particle calculation (e.g., Hartree-Fock, or DFT) for spin-channel $\sigma$. 
In the $GW$ approximation, the self-energy $\Sigma$ is given by
\begin{equation}\label{eq:sigma}
\Sigma_\sigma ({\bf r},{\bf r'},\omega) =
i\int \frac{d\omega'}{2\pi} G_\sigma({\bf r},{\bf r'},\omega+\omega')
W({\bf r},{\bf r'},\omega')e^{i\omega\eta} \quad.
\end{equation}
The screened Coulomb interaction $W$, in turn, 
is also determined from the solution of a Dyson-like equation 
\begin{align}
W&({\bf r},{\bf r'},\omega) = v({\bf r},{\bf r'}) + \nonumber \\ 
&\int d{\bf r}_1d{\bf r}_2  v({\bf r},{\bf r}_1) \chi_0({\bf r}_1,{\bf r}_2,\omega) W({\bf r}_2,{\bf r'},\omega)\quad,
\label{eq:W}
\end{align}
where $v({\bf r},{\bf r'})=|{\bf r}-{\bf r'}|^{-1}$  is the bare Coulomb interaction.
The polarizability $\chi_0$ is most easily expressed on the time axis $\tau$ 
\begin{align}\label{eq:chi}
\chi_0({\bf r},{\bf r'},\tau) = -i \sum_\sigma G_{\sigma}({\bf r},{\bf r'},\tau)G_{\sigma}({\bf r'},{\bf r},-\tau)\quad.
\end{align}
and is Fourier transformed to the frequency axis before it is used in Eq.~(\ref{eq:W}).

The structure of Eqs.~(\ref{eq:dyson})-(\ref{eq:chi}) reveals the self-consistent nature of the $GW$ approximation. 
Due to the interdependence of $G$, $\chi_0$, $W$, and $\Sigma$,  
Eqs.~(\ref{eq:dyson})-(\ref{eq:chi}) need to be solved iteratively 
until the satisfaction of a given convergence criterion.\cite{caruso/2013} 
We denote the procedure in which Eqs.~(\ref{eq:dyson})-(\ref{eq:chi})  are solved fully 
self-consistently as sc$GW$.
Recent studies have revealed that Hedin's equations  
may exhibit multiple-solution behaviour \cite{Lani2012,Lischner2012,Berger2014,Tandetzky2015,Stan2015,Scherpelz2016}. 
For closed shell molecules we have not yet observed multiple solutions. Moreover, it has been shown that, 
if self-consistency is achieved through the solution of the Dyson equation, as in this work, the 
self-consistent loop converges to the unique physical solution \cite{Stan2015}.

In sc$GW_0$ the screened interaction $W$ is evaluated only once using 
orbitals and eigenvalues from an independent-particle calculation.
The Dyson equation is thus solved iteratively updating $G$ and 
$\Sigma$  at each step, but keeping $W_0$ fixed.
In sc$GW$ and sc$GW_0$, the physical properties of the system 
-- such as, e.g., the total energy\cite{dahlen/2006,stan/2006,stan/2009,caruso/2013b,hellgren/2015}, 
the electron density,\cite{caruso/2014} and the ionization energy \cite{rostgaard/2010,caruso/2012} --
may be extracted directly from the self-consistent Green's function by means of the spectral function
\begin{equation}\label{eq:spectrum}
A(\omega)=-\frac{1}{\pi}\int d{\bf r}\lim_{{\bf r'}\rightarrow{\bf r}}{\rm Im} G({\bf r},{\bf r'},\omega) \quad .
\end{equation}
%
  \begin{figure}
  \includegraphics[width=0.68\textwidth]{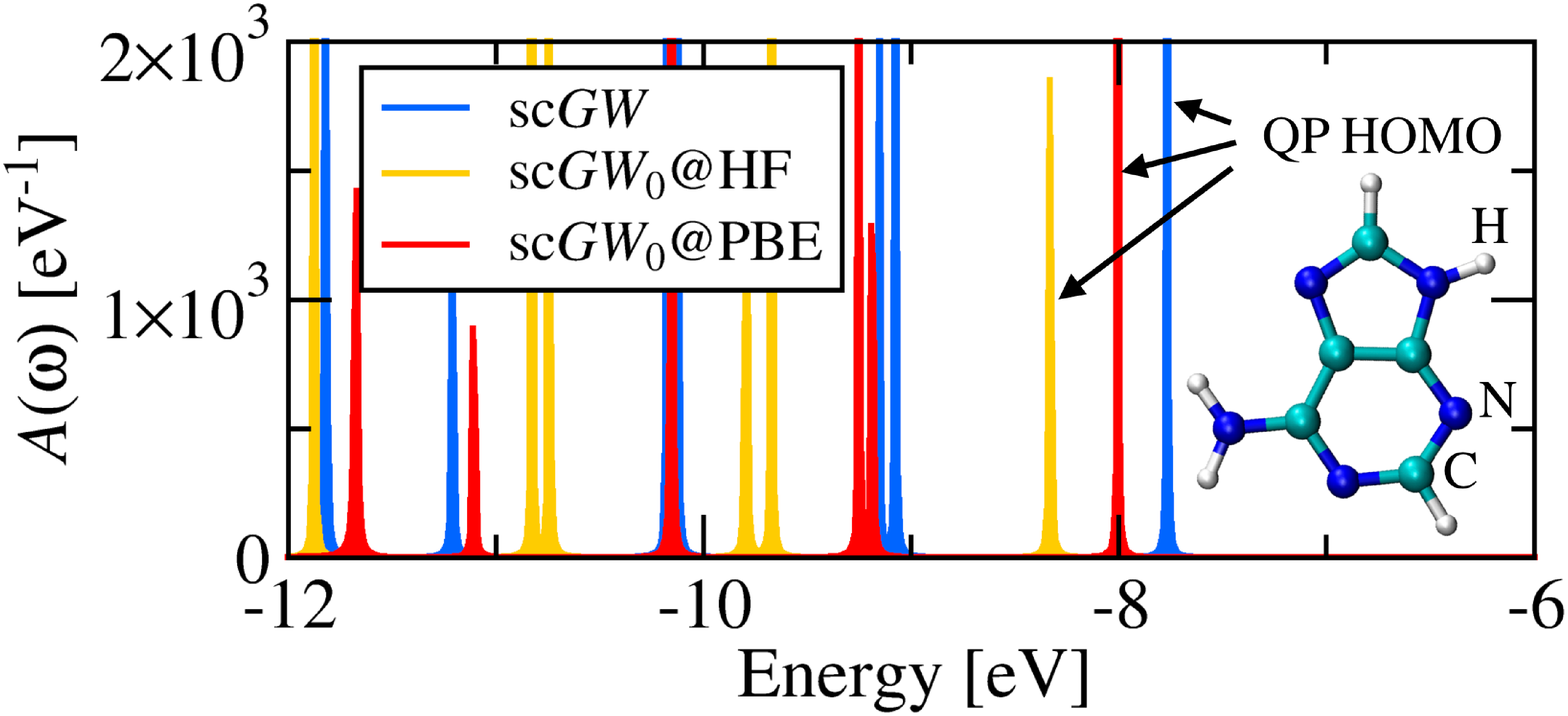}
  \caption{Spectral function of the adenine nucleobase, for which the molecular 
geometry in shown, obtained from sc$GW$, sc$GW_0$@HF, and sc$GW_0$@PBE using  
the def2-TZVPP basis set \cite{weigend/2005}. The quasiparticle HOMO is indicated by arrows. }
  \label{fig:spectrum}
  \end{figure}
As an example, we report in Fig.~\ref{fig:spectrum} the spectral function of the adenine {nucleobase} (C$_5$H$_5$N$_5$O)
evaluated using sc$GW$, sc$GW_0$@HF, and sc$GW_0$@PBE. For each approach, the energy of {the} quasiparticle HOMO 
is given by the position of {the} highest energy peak, indicated by arrows in Fig.~\ref{fig:spectrum}.
We note that sc$GW_0$ still exhibits a dependence on the starting point, which stems from the non-self-consistent 
treatment of $W$, whereas sc$GW$ {is} completely independent of the initial reference calculation.\cite{caruso/2012}  

In the $G_0W_0$ approach, the quasiparticle energies $\epsilon^{\rm QP}$ are evaluated as a first-order perturbative 
correction to a set of single-particle 
(SP) eigenvalues $\epsilon^{\rm SP}$ [obtained, for instance, from DFT] 
\begin{equation}\label{eq:QPE}
\epsilon^{\rm QP}_{n\sigma} = 
\epsilon^{\rm SP}_{n\sigma} + \langle \psi_n^\sigma | \Sigma(\epsilon^{\rm QP}_{n\sigma})- v_{\mathrm{0}}| \psi_n^\sigma\rangle\quad .
\end{equation}
Owing to the perturbative nature of Eq.~(\ref{eq:QPE}), one would expect a pronounced dependence of 
$\epsilon_{n\sigma}^{\rm QP}$ on the starting point, that is, on the set of eigenvalues $\epsilon_{n\sigma}^{\rm SP}$ and orbitals $\psi_{n\sigma}$. To benchmark the starting point dependence for the $GW$100 test set we consider hereafter 
two different starting points: Hartree-Fock and 
the Perdew-Burke-Ernzerhof\cite{PBE96} (PBE) generalized gradient approximation to DFT.
We explicitly denote the starting-point dependence by adopting the notation 
{\it method}$@${\it starting point} (e.g., $G_0W_0$@PBE).

In the qs$GW$ self-consistency treatment the Green's function keeps the analytic structure of a non-interacting Green's function (omitting spin indices for brevity)
\begin{align}
\label{qsgwgf}
G_0^{\rm{qs}GW}({\bf r},{\bf r'},\omega)=\sum_n \frac{\psi_{n}^{\rm QP}({\bf r})\psi_{n}^{*\rm QP}({\bf r'})}
{\omega-(\epsilon_{n}^{\rm QP}-\mu)-i\eta \,sgn(\mu-\epsilon_{n}^{\rm QP})}.
\end{align}
The quasi-particle orbitals and energies are iteratively updated solving the quasi-particle equation applying a linear mixing scheme.\cite{faleev/2004,vanschilfgaarde/2006,kaplan16} The QP-orbitals of the $(i+1)$th iteration $\psi^{(i+1)}_{n}({\bf r}) $ are expressed in terms of the orbitals of the previous iteration
\begin{equation} \label{eq:ks2qp}
\psi^{(i+1)}_{n}({\bf r}) = \sum_{\underline{n}}  \mathcal{A}^{(i+1)}_{n \underline{n}} \psi^{(i)}_{\underline{n}}({\bf r}).
\end{equation}
In the reference basis $\psi^{(i)}_{\underline{n}}({\bf r})$ Eq.~(\ref{eq:QPE}) takes the form of an eigenvalue problem
\begin{eqnarray}\label{eq:qpe2}
\sum_{\underline{n}} \mathcal{A}^{(i+1)}_{n' \underline{n}} \bigg [ \int \!\! \text{d}{\bf r}\:\text{d}{\bf r}'  \phantom{.}\psi^{(i)}_{n}({\bf r}) \big( H_0[G_0^{(i)}] \delta({\bf r}-{\bf r}') \nonumber\\ + \tilde{\Sigma}({\bf r}, {\bf r}') \big) \psi^{(i)}_{\underline{n}}({\bf r}')  \bigg ] = \epsilon^{{\rm QP}(i+1)}_{n'}\mathcal{A}^{(i+1)}_{n' n}
\end{eqnarray}
where $H_0[G_0^{\rm{qs}GW}]$ is the single-particle part of the Hamiltonian evaluated with the electron density generated by $G_0^{\rm{qs}GW}$. The self-energy matrix is approximated as static and Hermitian
\begin{equation}\label{eq:effSelf}
 \tilde{\Sigma}_{nn'} = \frac{1}{2} \left( \Sigma_{n n'}(\epsilon_{n}) +  \Sigma_{n n'}(\epsilon_{n'})\right) .
\end{equation}
The diagonalization of Eq.~(\ref{eq:qpe2}) updates $\epsilon^{{\rm QP}(i+1)}_{n'}$ and $\mathcal{A}^{(i+1)}_{n' n}$. With these new orbitals, the wave functions at iteration $i+1$ ($\psi^{{\rm QP}(i+1)}_{n}({\bf r})$) are constructed via Eq.~(\ref{eq:ks2qp}). 
The orbitals become orthonormal by construction due to the hermiticity of the operators in Eq.~(\ref{eq:qpe2}). 

qs$GW$ is closely related to $G_0W_0$ in the sense that in each cycle of the self-consistent solution the Green's function is a non-interacting $G_0$. The final result was shown to be independent of the starting point,\cite{kaplan16} 
but both the stability of the iterative cycle and the rate of convergence can be greatly improved by using an optimal starting point. In addition, it was found that a simple iteration scheme may not always converge. In practice linear mixing scheme is applied. In qs$GW$ the orbital energies are directly available via Eq.~(\ref{eq:qpe2}).

Beside sc$GW$, sc$GW_0$, and qs$GW$, other approximate self-consistent $GW$ approaches 
have been investigated in the past, such as eigenvalue self-consistent $GW$,\cite{blase/2011,faber/2011,marom/2012,kaplan15gwso} and $GW$+COHSEX.\cite{bruneval/2006,Knight/etal:2016} These will not be discussed in this article.

Among the different flavors of $GW$ calculations, the starting-point dependence 
is most pronounced in $G_0W_0$, since both $G_0$ and $W_0$ depend 
explicitly on the initial set of orbitals and eigenvalues.
Yet, this ambiguity also provides a {means} to improve the accuracy of $G_0W_0$, by 
seeking the optimal starting point that leads to the satisfaction of
exact physical constraints. A prominent example is the piecewise 
linearity of the total energy.\cite{PPLB82} Usually approximate 
theories do not automatically exhibit a linearly changing 
total energy under fractional electron removal (or addition) but 
instead produce a {DSLE}. 
If the total energy were a linear function of the fractional particle number, the ionization energy of 
the neutral system would be equal to the electron affinity of the 
cation (EA$_{\rm c}$).\cite{yang_jcp_lumo_2012,Atalla/etal:2016} Identifying the ionization energy with 
the $G_0W_0$ quasiparticle HOMO and EA$_{\rm c}$ with the LUMO 
of the cationic system, one may thus define the DSLE as\cite{dauth/2015}
\begin{equation}\label{eq:dsle}
 \Delta_{\rm DSLE} \equiv \epsilon^{\rm QP}_{\rm HOMO} - \epsilon^{\rm QP}_{\rm LUMO,\rm c}. 
\end{equation}
This definition can be applied to approximately quantify the DSLE in 
the $GW$ method without explicitly invoking the total energy at 
fractional particle numbers.
Furthermore, the minimization of $\Delta_{\rm DSLE}$ in $G_0W_0$ calculations 
allows one to find a starting point that minimizes or completely eliminates the DSLE.
We here adopt the DSLE-min $GW$ approach proposed in Ref.~\citenum{dauth/2015} which is based 
on these concepts.
For the DSLE-min procedure we utilize starting points from PBE-based hybrid (PBEh) 
functionals\cite{PBE0} with an adjustable fraction $\alpha$ of Hartree-Fock exchange and evaluate Eq.~(\ref{eq:dsle}) with the $G_0W_0$@PBEh$(\alpha)$ quasiparticle energies.
We then identify the optimal starting point with the very $\alpha$ that leads to a minimization of $\Delta_{\rm DSLE}$.

The coupled cluster singles, doubles, and perturbative triples [CCSD(T)]\cite{purvis82, raghavachari89, szabo/1989} approach is often regarded as the {\it gold standard} among the quantum chemistry methods as it yields results 
that approach chemical accuracy for a variety of physical/chemical properties, such as binding energies 
and atomization energies. 
CCSD(T) values are thus particularly suitable to {unambiguously} establish the 
accuracy of $GW$ approaches for the ionization energies. 
In the following, our calculated ionization energies are compared to reference values from CCSD(T),\cite{krause/2015} whereby 
the ionization energy has been obtained as a total energy difference between the ionized and neutral molecules. 
The comparison to CCSD(T) is here preferred to experimental data 
as it allows us to focus on the effects of exchange and correlation. 
We can therefore safely ignore the effects of temperature, nuclear vibrations, 
and interaction with the environment, which affect experimental ionization energies.\cite{gallandi/2015} 
The CCSD(T) calculations of Ref.~\citenum{krause/2015} used the 
molecular geometries of the $GW$100 test set\cite{vansetten/2015}, and are 
therefore suitable to be compared with our calculations, in which the same 
geometries were employed. 
Additional details on the CCSD(T) calculations may be found in Ref.~\citenum{krause/2015}.

\section{Computational details}

{
Our $G_0W_0$, DSLE-min $GW$, and sc$GW$ calculations have been performed with the 
{\tt FHI-aims} code \cite{blum/2009,HavuV09,ren/2012}, whereas qs$GW$ calculations 
have been performed using a local version of the TURBOMOLE \cite{TURBOMOLE} code. 
For $G_0W_0$, DSLE-min $GW$, and sc$GW$ the frequency dependence is treated
on the imaginary frequency axis and the quasiparticle energies are extracted 
by performing an analytic continuation based on Pad\'e approximants.
Similarly to Ref.~\citenum{vansetten/2015}, for $G_0W_0$ and DSLE-min $GW$ the 
parametrization of the analytic continuation employed 200 imaginary frequency 
points on a Gauss-Legendre grid and 16 poles
for the Pad\'e approximant method. 
The qs$GW$ calculations were performed directly in real frequency by exploiting 
the full analytic structure of $G$ and $W$ as described in Ref.~\citenum{vansetten13, kaplan16}.
Our sc$GW$ calculations used the same computational parameters as Ref.~\citenum{caruso/2012,caruso/2013} for the frequency dependence.
At variance with Ref.~\citenum{vansetten/2015}, no basis set extrapolation scheme has been employed in this work.}
Additional details on the numerical implementations of $G_0W_0$, 
sc$GW$, and sc$GW_0$ in {\tt FHI-aims}\cite{ren/2012,caruso/2012,caruso/2013} 
and the qs$GW$ implementation in TURBOMOLE\cite{vansetten13, kaplan16}
can be found elsewhere. 
All calculations use the same parameters 
reported in Ref.~\citenum{vansetten/2015} 
for the resolution-of-identity, and the real-space grids.
To {enable the direct} comparison with reference values from CCSD(T), 
we used the Gaussian def2-TZVPP basis sets.\cite{weigend/2005}
In FHI-aims the Gaussian basis function are numerically 
tabulated and are treated as numerical orbitals. 
We refer to Ref.~\citenum{vansetten/2015} 
for detailed convergence tests for this procedure.
For the DSLE-min method, basis set converged calculations for the quasiparticle energies 
have been performed using the Tier~4 basis sets augmented by 
Gaussian aug-cc-pV5Z basis functions {(Tier~4$^+$)} \cite{ren/2012}. 
To facilitate the comparison with CCSD(T), 
we also report DSLE-min quasiparticle energies 
obtained with def2-TZVPP basis sets. 

We use the same geometries as in Ref.~\citenum{vansetten/2015}.
\footnote{Experimental geometries have been employed whenever available,
otherwise molecular geometries are optimized within the PBE approximation for the exchange-correlation functional 
using the def2-QZVP basis set. More details on the strategy adopted for selecting the compounds of the $GW$100 set 
and their geometries are given in Ref.~\citenum{vansetten/2015}.}
We assume zero electronic temperature and the effects of nuclear vibrations are ignored. 
All ionization energies are vertical and do not include any relativistic corrections.

\section{Ionization energies for the \textit{GW}100 set}\label{sec:IE}

The $GW$100 test set consists of 100 atoms and molecules which have been selected to span a broad range 
of chemical bonding situations, chemical compositions, and ionization energies.
Due to the absence of all-electron def2-TZVPP basis sets for fifth-row elements, 
we exclude Xe, Rb$_2$, Ag$_2$, and the iodine-{containing} compounds (I$_2$, C$_2$H$_3$I, CI$_4$, and AlI$_3$). For the remaining 93 member of $GW$100 we can then conduct a meaningful comparison with CCSD(T) reference data.

{As discussed in Ref.~\citenum{vansetten/2015}, many molecules of the GW100 testset have 
positive LUMO energies (that is, negative electron affinities), which makes them unsuitable for a systematic assessment
of electron affinities since experimental data for such compounds is difficult to obtain. Moreover, CCSD(T) reference data
is presently also not available for the LUMOs in the GW100 testset \cite{krause/2015}.
For these reasons, we focus here on the first vertical ionization energy, for which experimental
and CCSD(T) reference data are available.
An assessment of GW methods for electron affinities may found in Ref.~\citenum{Knight/etal:2016}. }
In table~1, we report the ionization energies for this subset of $GW$100 calculated with
qs$GW$, sc$GW$, sc$GW_0$@HF, and sc$GW_0$@PBE and def2-TZVPP basis sets. For comparison, we also report the CCSD(T) ionization energies from Ref.~\citenum{krause/2015}.

\begin{center}\tiny
\begin{longtable}{rllrrrrrrr}
\caption{Vertical ionization energies for the $GW$100 test set calculated with qs$GW$, sc$GW$, sc$GW_0$@HF, 
sc$GW_0$@PBE, and DSLE-minimized $G_0W_0$ (DSLE-min) and def2-TZVPP basis sets.  
Basis set converged DSLE-minimized $G_0W_0$ calculations employ 
Tier 4 basis sets augmented by Gaussian aug-cc-pV5Z basis functions, denoted as 
DSLE-min (T4+).
For comparison, we also report CCSD(T) values from Ref.~\cite{krause/2015}. All values are in eV. } \\
\hline\hline
&Name&Formula&qs$GW$&sc$GW$&sc$GW_0$@HF&sc$GW_0$@PBE&DSLE-min&DSLE-min (T4+)&CCSD(T) \\
\hline
\endfirsthead
\hline \multicolumn{8}{l}{{\tablename\ \thetable{} continued}} \\
\hline
&Name&Formula&qs$GW$&sc$GW$&sc$GW_0$@HF&sc$GW_0$@PBE&DSLE-min&DSLE-min (T4+)&CCSD(T) \\
\hline
\endhead
\hline \multicolumn{8}{r}{{Continues on next page}} \\ \hline
\endfoot
\hline \hline
\endlastfoot
1&Helium&He&-24.43&-24.44&-24.47&-24.01&-&-&-24.51\\
2&Neon&Ne&-21.62&-21.40&-21.49&-20.84&-20.82&-20.22&-21.32\\
3&Argon &Ar&-15.53&-15.26&-15.50&-15.18&-15.34&-15.27&-15.54\\
4&Krypton&Kr&-13.74&-13.65&-13.88&-13.62&-13.62&-13.75&-13.94\\
6&Hydrogen&H$_2$&-16.22&-16.18&-16.27&-15.98&-&-&-16.40\\
7&Lithium~dimer&Li$_2$&-5.34&-4.96&-5.15&-5.02&-5.00&-5.05&-5.27\\
8&Sodium~dimer&Na$_2$&-5.02&-4.63&-4.80&-4.74&-4.87&-4.93&-4.95\\
9&Sodium~tetramer&Na$_4$&-4.25&-3.85&-4.09&-3.99&-4.18&-4.27&-4.23\\
10&Sodium hexamer&Na$_6$&-4.41&-3.94&-4.25&-4.16&-4.31&-4.40&-4.35\\
11&Dipotassium&K$_2$&-4.08&-3.73&-3.90&-3.86&-3.96&-4.10&-4.06\\
13&Nitrogen&N$_2$&-16.01&-15.44&-15.84&-15.32&-15.49&-15.75&-15.57\\
14&Phosphorus~dimer&P$_2$&-10.40&-9.73&-10.20&-10.01&-10.30&-10.52&-10.47\\
15&Arsenic~dimer&As$_2$&-9.62&-9.00&-9.48&-9.34&-9.52&-9.82&-9.78\\\hline
16&Fluorine&F$_2$&-16.33&-15.78&-16.17&-15.50&-15.56&-15.76&-15.71\\
17&Chlorine&Cl$_2$&-11.52&-11.07&-11.47&-11.13&-11.36&-11.55&-11.41\\
18&Bromine&Br$_2$&-10.54&-10.23&-10.58&-10.30&-10.31&-10.77&-10.54\\
20&Methane &CH$_4$&-14.56&-14.28&-14.50&-14.14&-14.20&-14.35&-14.37\\
21&Ethane&C$_2$H$_6$&-12.99&-12.62&-12.92&-12.55&-12.60&-12.75&-13.04\\
22&Propane&C$_3$H$_8$&-12.35&-11.95&-12.30&-11.92&-12.03&-12.18&-12.05\\
23&Butane&C$_4$H$_{10}$&-11.89&-11.46&-11.85&-11.46&-11.73&-11.88&-11.57\\\hline
24&Ethylene&C$_2$H$_4$&-10.63&-10.14&-10.45&-10.24&-10.40&-10.60&-10.67\\
25&Acetylene&C$_2$H$_2$&-11.53&-10.89&-11.23&-10.98&-11.17&-11.43&-11.42\\\hline
26&tetracarbon&C$_4$&-11.45&-10.68&-11.21&-10.87&-10.87&-11.07&-11.26\\
27&Cyclopropane&C$_3$H$_6$&-11.13&-10.62&-10.98&-10.66&-10.77&-11.00&-10.87\\
28&Benzene&C$_6$H$_6$&-9.38&-8.73&-9.20&-8.97&-9.12&-9.34&-9.29\\
29&Cyclooctatetraene&C$_8$H$_8$&-9.30&-7.81&-8.33&-8.04&-8.21&-8.44&-8.35\\
30&Cyclopentadiene&C$_5$H$_6$&-8.73&-8.10&-8.54&-8.29&-8.47&-8.69&-8.68\\\hline
31&Vynil~fluoride&C$_2$H$_3$F&-10.64&-10.11&-10.46&-10.16&-10.36&-10.59&-10.55\\
32&Vynil~chloride&C$_2$H$_3$Cl&-10.09&-9.63&-10.02&-9.72&-9.92&-10.14&-10.09\\
33&Vynil~bromide&C$_2$H$_3$Br&-9.33&-8.82&-9.19&-8.94&-9.06&-9.32&-9.27\\
35&Carbon~tetrafluoride&CF$_4$&-16.77&-16.34&-16.75&-15.89&-15.84&-15.78&-16.30\\
36&Carbon~tetrachloride&CCl$_4$&-11.63&-11.16&-11.69&-11.18&-11.46&-11.57&-11.56\\
37&Carbon~tetrabromide&CBr$_4$&-10.57&-10.10&-10.59&-10.16&-10.33&-10.59&-10.46\\
39&Silane&SiH$_4$&-13.04&-12.74&-13.00&-12.55&-12.66&-12.88&-12.80\\
40&Germane&GeH$_4$&-12.81&-12.40&-12.67&-12.28&-12.41&-12.55&-12.50\\
41&Disilane&Si$_2$H$_6$&-10.88&-10.46&-10.82&-10.45&-10.48&-10.75&-10.65\\
42&Pentasilane&Si$_5$H$_{12}$&-9.56&-9.04&-9.50&-9.10&-9.18&-9.32&-9.27\\\hline
43&Lithium~hydride&LiH&-8.00&-7.88&-7.97&-7.45&-6.48&-6.71&-7.96\\
44&Potassium~hydride&KH&-6.17&-6.02&-6.17&-5.52&-5.65&-5.63&-6.13\\
45&Borane&BH$_3$&-13.52&-13.22&-13.42&-13.05&-13.17&-13.30&-13.28\\
46&Diborane(6)&B$_2$H$_6$&-12.58&-12.23&-12.54&-12.09&-12.17&-12.30&-12.26\\
47&Ammonia&NH$_3$&-11.08&-10.76&-10.97&-10.59&-10.59&-10.78&-10.81\\
48&Hydrogen~azide&HN$_3$&-10.91&-10.24&-10.69&-10.38&-10.61&-10.89&-10.68\\
49&Phosphine&PH$_3$&-10.65&-10.24&-10.53&-10.28&-10.39&-10.60&-10.52\\
50&Arsine&AsH$_3$&-10.50&-10.10&-10.39&-10.17&-10.24&-10.49&-10.40\\
51&Hydrogen~sulfide&SH$_2$&-10.39&-9.97&-10.26&-10.02&-10.15&-10.38&-10.31\\
52&Hydrogen~fluoride&FH&-16.33&-16.11&-16.26&-15.71&-15.63&-15.64&-16.03\\
53&Hydrogen~chloride&ClH&-12.65&-12.28&-12.55&-12.27&-12.41&-12.57&-12.59\\\hline
54&Lithium~fluoride&LiF&-11.52&-11.34&-11.50&-10.59&-10.66&-10.85&-11.32\\
55&Magnesium~fluoride&F$_2$Mg&-13.99&-13.77&-13.97&-13.05&-12.87&-13.00&-13.71\\
56&Titanium~fluoride&TiF$_4$&-15.75&-15.55&-16.15&-14.98&-14.80&-15.19&-15.48\\
57&Aluminum~fluoride&AlF$_3$&-15.69&-15.40&-15.68&-14.83&-14.59&-14.75&-15.46\\
58&Fluoroborane &BF&-11.13&-10.64&-10.94&-10.56&-10.82&-10.98&-11.09\\
59&Sulfur~tetrafluoride&SF$_4$&-12.98&-12.47&-12.95&-12.36&-12.51&-12.73&-12.59\\
60&Potassium~bromide&BrK&-8.15&-7.88&-8.12&-7.72&-7.84&-8.25&-8.13\\
61&Gallium~monochloride&GaCl&-9.80&-9.35&-9.69&-9.49&-9.62&-9.93&-9.77\\
62&Sodium~chloride&NaCl&-9.07&-8.79&-9.03&-8.51&-8.79&-9.03&-9.03\\
63&Magnesium~chloride&MgCl$_2$&-11.64&-11.39&-11.70&-11.24&-11.22&-11.39&-11.67\\
65&Boron~nitride&BN&-11.79&-11.06&-11.58&-11.27&-11.19&-11.81&-11.89\\
66&Hydrogen~cyanide&NCH&-13.65&-13.15&-13.51&-13.20&-13.47&-13.72&-13.87\\
67&Phosphorus~mononitride&PN&-11.93&-11.56&-12.03&-11.60&-11.60&-11.84&-11.74\\
68&Hydrazine&H$_2$NNH$_2$&-10.08&-9.63&-9.93&-9.52&-9.53&-9.75&-9.72\\\hline
69&Formaldehyde&H$_2$CO&-11.22&-10.82&-11.15&-10.67&-10.77&-11.02&-10.84\\
70&Methanol&CH$_4$O&-11.46&-11.07&-11.36&-10.86&-10.94&-11.19&-11.04\\
71&Ethanol&C$_2$H$_6$O&-11.07&-10.69&-11.05&-10.51&-10.59&-10.84&-10.69\\
72&Acetaldehyde&C$_2$H$_4$O&-10.62&-10.20&-10.59&-10.03&-10.10&-10.36&-10.21\\
73&Ethoxy~ethane&C$_4$H$_{10}$O&-10.23&-9.81&-10.27&-9.67&-9.77&-10.02&-9.82\\
74&formic~acid&CH$_2$O$_2$&-11.78&-11.42&-11.80&-11.19&-11.29&-11.57&-11.42\\\hline
75&Hydrogen~peroxide&HOOH&-11.98&-11.55&-11.90&-11.38&-11.42&-11.69&-11.59\\
76&Water&H$_2$O&-12.91&-12.59&-12.78&-12.32&-12.26&-12.45&-12.57\\
77&Carbon~dioxide&CO$_2$&-14.07&-13.55&-13.95&-13.45&-13.61&-13.91&-13.71\\
78&Carbon~disulfide&CS$_2$&-10.04&-9.45&-9.95&-9.69&-9.89&-10.14&-9.98\\
79&Carbon~oxysulfide&OCS&-11.33&-10.72&-11.17&-10.88&-11.08&-11.35&-11.17\\
80&Carbon~oxyselenide&OCSe&-10.60&-10.00&-10.42&-10.19&-10.29&-10.62&-10.79\\
81&Carbon~monoxide&CO&-14.55&-13.95&-14.43&-13.90&-14.21&-14.44&-14.21\\
82&Ozone&O$_3$&-13.21&-12.54&-13.16&-12.57&-12.24&-12.49&-12.55\\
83&Sulfur~dioxide&SO$_2$&-12.54&-12.05&-12.54&-12.06&-12.21&-12.55&-13.49\\
84&Beryllium~monoxide&BeO&-10.11&-9.77&-10.01&-9.58&-9.40&-9.68&-9.94\\
85&Magnesium~monoxide&MgO&-8.30&-7.97&-8.27&-7.72&-7.48&-7.60&-7.49\\\hline
86&Toluene&C$_7$H$_8$&-9.00&-8.35&-8.83&-8.60&-8.74&-8.96&-8.90\\
87&Ethylbenzene&C$_8$H$_{10}$&-8.97&-8.30&-8.80&-8.55&-8.68&-8.91&-8.85\\
88&Hexafluorobenzene&C$_6$F$_6$&-9.91&-9.48&-10.08&-9.66&-9.96&-10.23&-9.93\\
89&Phenol&C$_6$H$_5$OH&-8.82&-8.19&-8.67&-8.39&-8.52&-8.78&-8.70\\\hline
90&Aniline&C$_6$H$_5$NH$_2$&-8.12&-7.51&-7.99&-7.69&-7.83&-8.09&-7.99\\
91&Pyridine&C$_5$H$_5$N&-9.76&-9.11&-9.58&-9.37&-9.53&-9.76&-9.66\\
92&Guanine&C$_5$H$_5$N$_5$O&-7.95&-7.49&-8.06&-7.71&-7.88&-8.18&-8.03\\
93&Adenine&C$_5$H$_5$N$_5$O&-8.41&-7.77&-8.33&-8.00&-8.16&-8.45&-8.33\\
94&Cytosine&C$_4$H$_5$N$_3$O&-8.99&-8.38&-8.93&-8.47&-8.63&-8.92&-9.51\\
95&Thymine&C$_5$H$_6$N$_2$O$_2$&-9.30&-8.69&-9.25&-8.83&-9.01&-9.28&-9.08\\
96&Uracil&C$_4$H$_4$N$_2$O$_2$&-9.74&-9.12&-9.66&-9.22&-9.41&-9.69&-10.13\\
97&Urea&CH$_4$N$_2$O&-10.45&-10.02&-10.45&-9.81&-10.13&-10.44&-10.05\\\hline
99&Copper~dimer&Cu$_2$&-7.52&-6.98&-7.23&-7.29&-7.15&-7.57&-7.57\\
100&Copper~cyanide&NCCu&-10.97&-10.54&-11.13&-10.26&-10.38&-10.50&-10.85\\
\hline
\end{longtable}
\end{center}

  \begin{figure}[t]
  \includegraphics[width=0.68\textwidth]{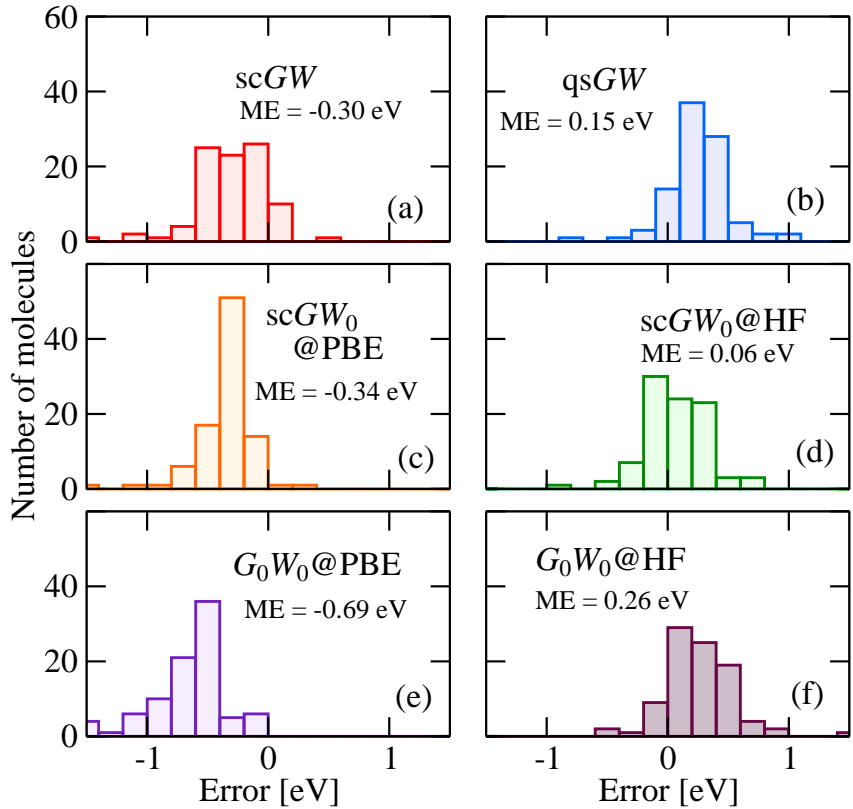}
  \caption{
            Error distribution [defined as the difference to CCSD(T) reference energies from Ref.~\citenum{krause/2015}] 
		for the ionization energies of the $GW$100 test set evaluated using 
(a) sc$GW$,
(b) qs$GW$,
(c) sc$GW_0$@PBE,
(d) sc$GW_0$@HF,
(e) $G_0W_0$@PBE, and
(f) $G_0W_0$@HF and def2-TZVPP basis sets.
The mean error (ME) for each method is listed in the corresponding panel.}
  \label{fig:error}
  \end{figure}

  \begin{figure}[t]
  \includegraphics[width=0.68\textwidth]{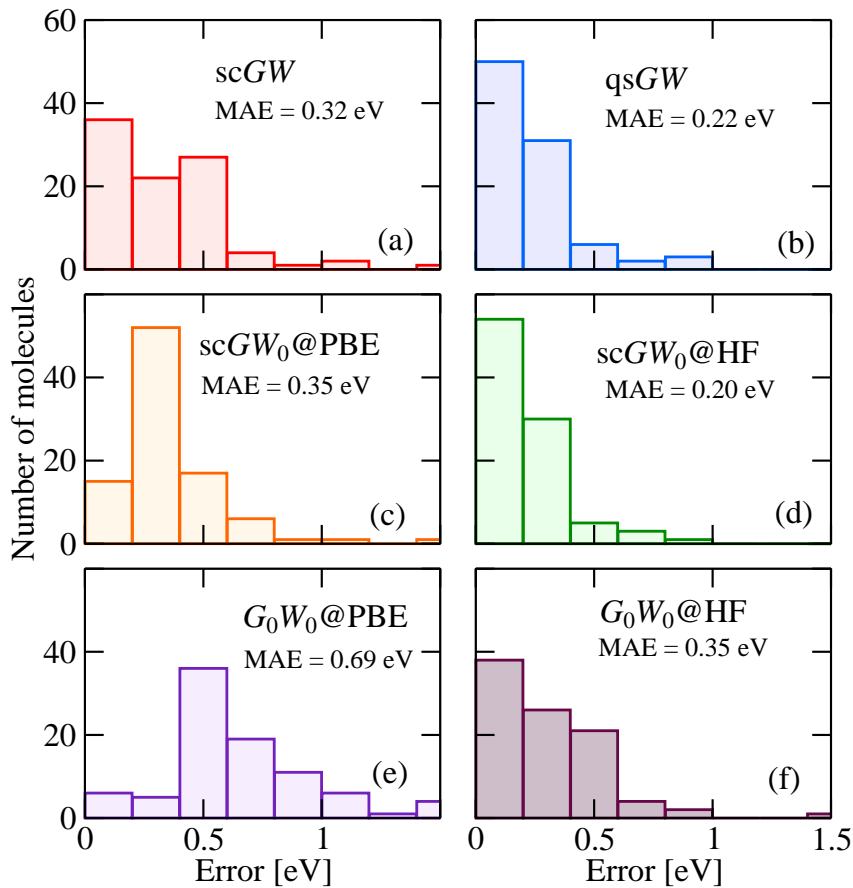}
  \caption{
  	          Absolute error distribution (defined similarly to Fig.~\ref{fig:error})
                for the ionization energies of the $GW$100 test set evaluated using 
(a) sc$GW$, 
(b) qs$GW$,
(c) sc$GW_0$@PBE,
(d) sc$GW_0$@HF,  
(e) $G_0W_0$@PBE, and
(f) $G_0W_0$@HF and def2-TZVPP basis sets. 
The mean absolute error (MAE) for each method is listed in the corresponding panel.}
  \label{fig:abserror}
  \end{figure}
  
  \begin{figure*}
  \includegraphics[width=0.9\textwidth]{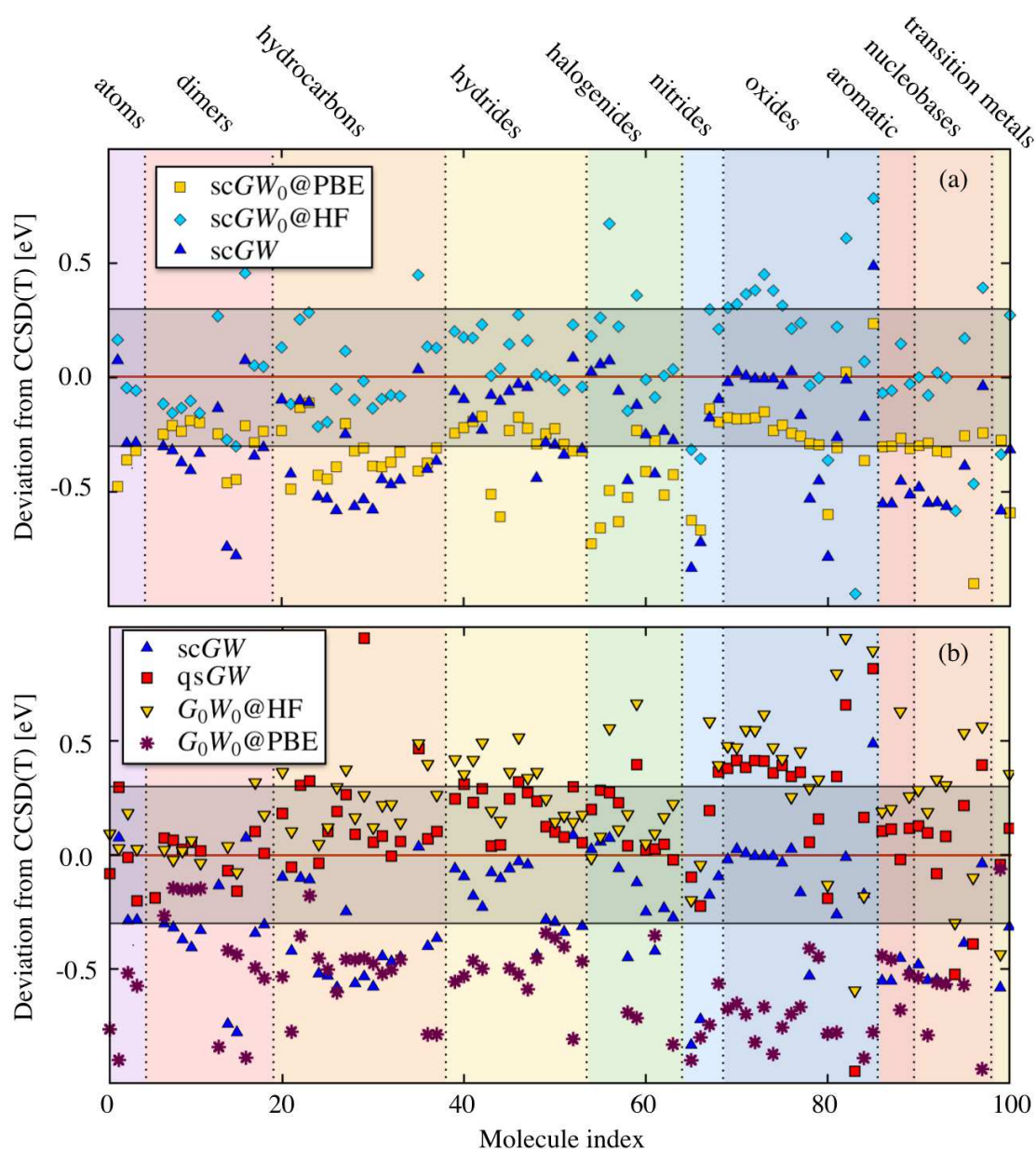}
  \caption{
Deviation between the CCSD(T) reference ionization energies and our first-principles calculations obtained using 
(a) sc$GW$, 
sc$GW_0$@PBE, and
sc$GW_0$@HF, 
and (b) sc$GW$, qs$GW$, $G_0W_0$@PBE, and
$G_0W_0$@HF and def2-TZVPP basis sets. 
Only compounds with ionization energies that differ from CCSD(T) by less than 1~eV are included.
Vertical dotted lines denotes the separation between different subgroups of the $GW$100 test set
and coincide with the horizontal separation lines of table~1. 
The separation in subgroups (as well as the name attributed to each subgroup) 
is a guide to the eye, but not necessarily representative of the chemical compositions of each compound. 
Points falling within the horizontal shaded area differ by less the 0.3~eV from CCSD(T).
}   \label{fig:deviation}
  \end{figure*}

\section{Comparison of \textit{GW} Methods}\label{sec:g0w0vsSC}
To quantify the deviation from CCSD(T) calculations, we analyse 
the error $\Delta\equiv \varepsilon^{\rm HOMO}_{\rm CCSD(T)} - \varepsilon^{\rm HOMO}_{\rm QP}$ 
and the  absolute error $\Delta_{\rm abs}\equiv |\varepsilon^{\rm HOMO}_{\rm CCSD(T)} - \varepsilon^{\rm HOMO}_{\rm QP}|$.
In Fig.~\ref{fig:error}, we report the error distribution for the 
molecules of the $GW$100 test set, whereas the absolute error is 
reported in Fig.~\ref{fig:abserror}. 


\subsection{sc\textit{GW} vs qs\textit{GW}}
\label{sec:scGWvsqsGW}

We start by considering the sc$GW$  and qs$GW$ approaches. At variance with $G_0W_0$ and sc$GW_0$, the sc$GW$
ionization energies are independent of the starting point.\cite{caruso/2012,caruso/2013} Any deviations between sc$GW$ and CCSD(T) can then be attributed to intrinsic limitations of the $GW$ approximation (i.e. missing vertex corrections) rather than the artificial starting-point dependence introduced by perturbation theory or approximate self-consistent procedures. The qs$GW$ ionization energies of molecules have also been reported to be independent of the starting point.\cite{kaplan16} However, for some solids, a dependence on the starting point has been observed \cite{carter2011}.

Our calculations reveal that qs$GW$ overestimates the ionization potentials in our test set by 0.15~eV 
on average [Fig.~\ref{fig:error}~(b)], whereas sc$GW$ underestimates them by 0.3~eV [Fig.~\ref{fig:error}~(a)].
qs$GW$ exhibits a MAE of $\sim0.22$~eV [Fig.~\ref{fig:abserror}~(b)] and it thus
yields quasiparticle energies in slightly better agreement with CCSD(T) than sc$GW$ [MAE$=0.32$~eV, Fig.~\ref{fig:abserror}~(a)].
Overall,  sc$GW$ and qs$GW$
ionization energies  differ on average by 0.45 eV, revealing that different forms of self-consistency
may affect significantly the value of the quasiparticle energies and the corresponding agreement with 
experiment. In the following we explore four different potential explanations. 

\subsubsection{Screening properties}

While it is expected that different forms of self-consistency lead to different results, the magnitude of the difference is surprising. At first glance, sc$GW$ and qs$GW$ should be similar since in both approaches the quasiparticle energies enter the denominator of the Green's function. For both approaches we would therefore expect underscreening, due to the inverse dependence of the magnitude of screening on the  energy difference between the lowest unoccupied and the highest occupied state in $GW$. In a beyond-$GW$ treatment this underscreening due to the large quasiparticle gap would be compensated by vertex corrections, such as ladder diagrams. \cite{Bruneval2005,Shiskin/etal:2007} Without this compensation,  the underscreening due to the too large quasiparticle gap in $W$ would lead to an overestimation of ionization energies and quasiparticle energies that resemble those of $G_0W_0$@HF, which is also based on an underscreened $W_0$ due to the large HOMO-LUMO gap in HF. For qs$GW$ we indeed observe this resemblance with $G_0W_0$@HF in Fig.~\ref{fig:error} and \ref{fig:abserror}, which results in the aforementioned slight average overestimation of ionization energies compared to CCSD(T). The small reduction of the ionization energies by 0.09~eV in going from $G_0W_0$@HF to qs$GW$ can therefore be attributed to a reduction of the underscreening due to the fact that the qs$GW$ gap is smaller than the HF gap and to density changes that we will discuss in the following.

The corresponding ionization-energy histogram for sc$GW$ is closer to sc$GW_0$@PBE and $G_0W_0$@PBE than to $G_0W_0$@HF, with a concomitant underestimation of the CCSD(T) reference data. This observation is consistent with previous sc$GW$ calculations for molecules \cite{caruso/2012,marom/2012,caruso/2013,Koval/etal:2014,caruso/2014,Pinheiro/etal:2015,Knight/etal:2016} that observed a similar underestimation of the ionization potential. Also in sc$GW$ the HOMO-LUMO gaps is smaller than in $G_0W_0$@HF and smaller than in qs$GW$. sc$GW$ therefore underscreens less than qs$GW$ and we attribute part of the 0.45~eV average deviation between qs$GW$ and sc$GW$ to this difference in screening.

\subsubsection{Spectral-weight transfer}

For solids, a spectral-weight transfer from the main quasiparticle peaks to satellites has been reported for sc$GW$ calculations of the homogeneous electron gas. \cite{holm1998a} Schematically, the self-consistent Green's function can be written as $G=ZG_{qp}+\bar{G}$, where $Z$ is the spectral weight of the quasiparticle peak $G_{qp}$ and $\bar{G}$ the incoherent part of the spectral function. In qs$GW$ $Z$ is equal to one and $\bar{G}$ is zero.\cite{Kotani/etal_2:2007,kaplan16} Conversely, for sc$GW$ $Z$ is smaller than one and $\bar{G}$ larger than zero, as spectral weight is transferred from $G_{qp}$ to $\bar{G}$. This spectral weight transfer leads to an additional underscreening and an overestimation of band gaps in solids. \cite{eguiluz1998,kutepov,Chu/etal:2016} 

For small molecules there are no continuum states or collective excitations that could be excited at valence energies.\cite{caruso/2013b} The sc$GW$ spectral functions therefore are sharply peaked around the quasiparticle energies and the spectrum exhibits no signature of an incoherent background in the valence energy region \cite{caruso/2013b} as show in Fig.~\ref{fig:spectrum}. We would thus not expect any additional underscreening due to spectral-weight transfer, because $Z$ is equal to one and $\bar{G}$ is zero, just as for qs$GW$. The spectral-weight transfer concept can therefore not explain the consistent underestimation observed for molecules in sc$GW$. \cite{caruso/2012,marom/2012,caruso/2013,caruso/2014,Pinheiro/etal:2015,Knight/etal:2016}

\subsubsection{Self-consistent density}
  \begin{figure*}[t]
  \includegraphics[width=0.85\textwidth]{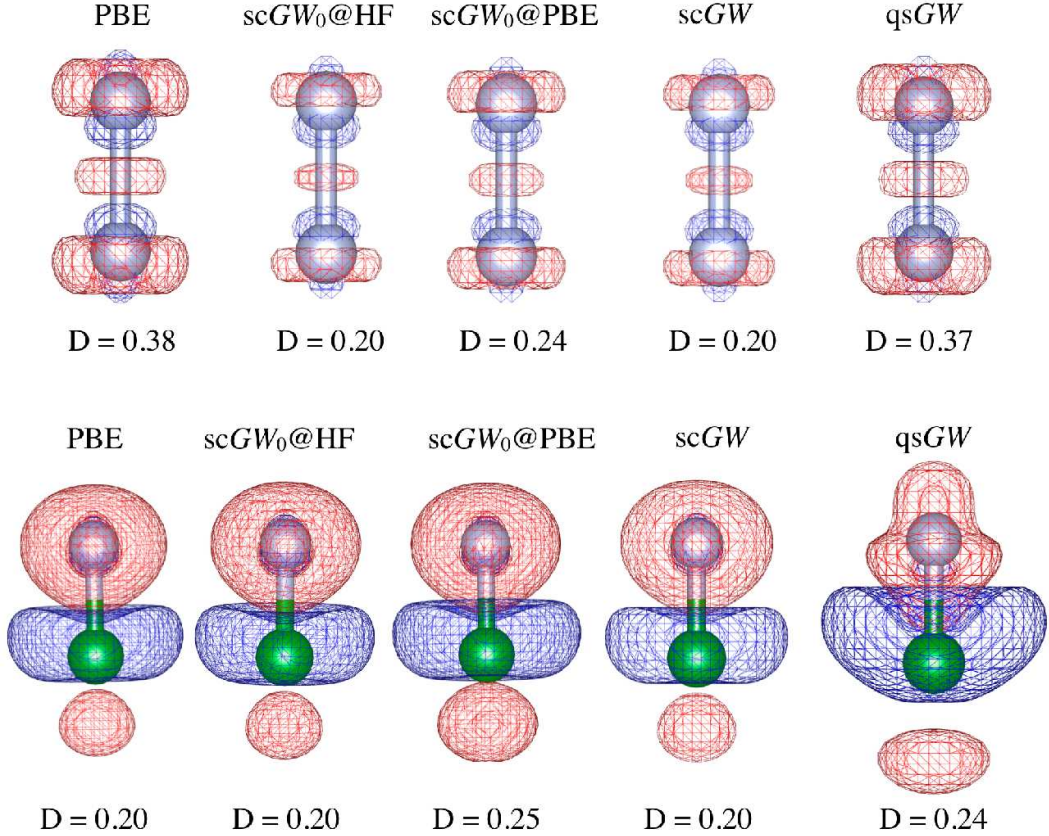}
  \caption{
  Isosurfaces of the density difference to Hartree-Fock for PBE, sc$GW_0$@HF, sc$GW_0$@PBE, sc$GW$, and qs$GW$. 
  We used an isovalue of 0.05 and 0.01~\AA$^{-3}$  for F$_2$ (upper panel) and BF (lower panel), respectively.} 
  \label{fig:dens}
  \end{figure*}

Further insight into the effects of different $GW$ approaches on electron correlation may be gained 
from the study of the self-consistent electron density. 
To focus on the effects of correlation, we consider in the following differences of  
the PBE, sc$GW$, sc$GW_0$, and qs$GW$ electron density to the density of a Hartree-Fock 
calculation using with the same computational parameters.
Figure~\ref{fig:dens} illustrates isosurfaces of these density 
differences for F$_2$ (upper panel) and BF (lower panel) with isovalues of 0.05 and 
0.01~\AA$^{-3}$, respectively.
To quantify the difference between the $GW$ and the HF densities, 
we introduce a density difference parameter $D$ defined as: 
\begin{align}
D= \int d{\bf r} \left| n^{GW}({\bf r}) - n^{\rm HF}({\bf r}) \right| 
\end{align}
for which the values for BF and F$_2$ are also reported in Fig.~\ref{fig:dens}.

For both BF and F$_2$, sc$GW$ and sc$GW_0$ induce qualitatively similar 
modifications of the electron density as compared to the 
Hartree-Fock reference both in shape and magnitude (as quantified by $D$).
In particular, both sc$GW$ and sc$GW_0$@HF yield $D=0.20$ for BF and F$_2$, whereas 
sc$GW_0$@PBE yields a slight larger modification of the electron density, quantified 
from the larger $D$ value, which we attribute to over-screening induced by the 
PBE starting point. In qs$GW$ the change of electron density is more pronounced with respect to sc$GW$ 
and, for the BF dimer, exhibits a considerably different charge redistribution pattern. 

Overall, these results indicate that electron densities resulting from sc$GW$ and qs$GW$ calculation may exhibit quantitative and qualitative differences. In self-consistent treatments, such density difference affect the external and the Hartree potential as well as the kinetic and the self-energy and thus contribute to the quasiparticle energy difference observed in this work. However, the small example shown in Fig.~\ref{fig:dens} illustrates that the density difference between qs$GW$ and sc$GW$ is neither systematic in shape nor in magnitude and can probably not explain the systematic shift of $\sim$0.45 eV observed between our qs$GW$ and sc$GW$ data.

\subsubsection{Kinetic energy}

{Another aspect in which sc$GW$ and qs$GW$ differ is the treatment of the kinetic energy. 
In the $G_0W_0$ approach, the quasiparticles are subject to the non-interacting kinetic energy.  
If the non-interacting Green's function $G_0$ derives, for example, from a Kohn-Sham DFT calculation
the kinetic energy contribution to the total energy would be that of 
the {\it fictitious} non-interacting system of Kohn-Sham particles ($T_s$).
In Kohn-Sham theory, the difference between $T_s$ and the kinetic energy of 
the interacting system $T$ -- as obtained for instance from a self-consistent Green's function 
calculation -- is included through the exchange-correlation energy functional. 
In the following, we analyze how the kinetic energy is handled in qs$GW$, a hybrid approach 
which combines elements of Green's theory and Kohn-Sham theory.
In particular, we discuss whether the differences in the sc$GW$ 
and qs$GW$ quasiparticle energies may be ascribed to a different 
treatment of the kinetic energy in the two methods.}



The difference between the non-interacting and the interacting kinetic energy of a $GW$ calculation
may be quantified by invoking the analogy with the random-phase approximation (RPA).\cite{Langreth1977,RPAreview} 
The total energy in sc$GW$, $G_0W_0$, and RPA can be separated into different contributions\cite{caruso/2013b,hellgren/2015}:
\begin{align}
E^{GW}[G]&=T[G]+E_{\rm ext}[G]+E_{\rm H}[G]+E_{\rm x}[G]+U^{GW}_{\rm c}[G] \\ \label{eq:E_GW}
E^{G_0W_0}[G_0]&=T_s[G_0]+E_{\rm ext}[G_0]+E_{\rm H}[G_0]+E_{\rm x}[G_0]+U^{GW}_{\rm c}[G_0] \\
E^{\rm RPA}[G_0]&=T_s[G_0]+E_{\rm ext}[G_0]+E_{\rm H}[G_0]+E_{\rm xc}^{\rm RPA}[G_0] \\
                            &=T_s[G_0]+E_{\rm ext}[G_0]+E_{\rm H}[G_0]+E_{\rm x}[G_0]+E^{\rm RPA}_{\rm c}[G_0]+T^{\rm RPA}_c[G_0]
\end{align}
where $T$ is the fully interacting kinetic energy, $T_s$ the non-interacting kinetic energy, 
$E_{\rm ext}$ the external energy, $E_{\rm H}$ the Hartree energy and $E_{\rm x}$ the exchange 
energy evaluated for the fully interacting Green's function $G$ or the non-interacting reference 
calculation $G_0$. 
Following Ref.~\citenum{caruso/2013b,hellgren/2015}, we defined 
$U_{\rm c}^{GW}$ and $E_{\rm c}^{\rm RPA}$ as the correlation energy functionals in the $GW$ and RPA approximation, respectively: 
\begin{align}
U_{\rm c}^{GW}&=\int_0^\infty \frac{d\omega}{2\pi}{\rm Tr}\{v[\chi(i\omega)-\chi_0(i\omega)]\}\label{eq-uc}\\
E_{\rm c}^{\rm RPA}&=\int_0^\infty \frac{d\omega}{2\pi}  \int_0^1\! d\lambda\,{\rm Tr}\{v[\chi_\lambda(i\omega)-\chi_0(i\omega)]\} \label{eq-Ec}
\end{align}
where $\chi_{\lambda}$ is the reducible polarizability
\begin{equation}
   \chi_{\lambda}=\chi_0 + \chi_0 v \chi_{\lambda}
\end{equation}
at coupling strength $\lambda$ that follows from the irreducible polarizability $\chi_0$ defined 
in Eq.~(\ref{eq:chi}). In $GW$ there is no coupling strength integration and $\chi=\chi_{\lambda=1}$. 
RPA contains a coupling strength integration over fictitious systems with coupling strength $\lambda$ 
that varies between zero (non-interacting) and one (fully interacting). The comparison between 
Eq.~(\ref{eq-uc}) and (\ref{eq-Ec}) reveals that $U_{\rm c}^{GW}$ contains only electronic correlation 
(that is, arising from the Coulomb interaction), whereas $E_{\rm c}^{\rm RPA}$ recaptures a interacting kinetic 
energy contribution through the coupling constant integration for the same starting point $G_0$.\cite{caruso/2013b}
We can then define the kinetic energy contribution of the correlation energy as
\begin{equation}
\label{eq-kincorr}
T_{\rm c}^{\rm RPA}[G_0]\equiv E^{\rm RPA}_{\rm c}[G_0]-U^{GW}_{\rm c}[G_0].
\end{equation}
Equations~(\ref{eq:E_GW}) to (\ref{eq-kincorr}) illustrate that both the full Green's function 
framework (sc$GW$) and DFT (e.g., RPA) incorporate the interacting kinetic energy. 
In the perturbative $G_0W_0$ framework, however, this contribution is absent. 

In sc$GW$ the quasiparticle energies are extracted directly from the imaginary part of the 
Green's function, i.e. the spectral function, as illustrated in Section~\ref{sec:method}, 
and therefore contain a contribution from the interacting kinetic energy. In DFT, the Kohn-Sham 
eigenvalues are obtained from the solution of the Kohn-Sham equation. The effective Kohn-Sham 
potential includes the exchange-correlation potential, that is defined as the functional 
derivative of the exchange-correlation energy $\frac{\delta E_{xc}}{\delta n}$ and therefore 
includes the difference between the interacting and the non-interacting kinetic energy 
in the correlation potential via the derivative of $T_c$.

Conversely, in the $G_0W_0$ approach, the quasiparticle energies $\epsilon^{\rm QP}$ are 
evaluated as a first-order perturbative correction to the single-particle eigenvalues 
$\epsilon^{\rm SP}$ as shown in Eq.~(\ref{eq:QPE}), which we repeat here for clarity
\begin{equation}\label{eq:QPE2}
\epsilon^{\rm QP}_{n\sigma} = 
\epsilon^{\rm SP}_{n\sigma} + \langle \psi_n^\sigma | \Sigma^{G_0W_0}(\epsilon^{\rm QP}_{n\sigma})- v_{\rm xc}| \psi_n^\sigma\rangle\quad .
\end{equation}
For DFT starting points, the matrix element of the exchange-correlation potential $v_{\rm xc}$ 
subtracts the aforementioned $T_c$ contribution from the eigenvalue $\epsilon^{\rm SP}_{n\sigma}$. 
Since $\Sigma^{G_0W_0}$ is purely an exchange and Coulomb correlation self-energy, it does not 
add an interacting kinetic energy contribution back in, which is thus absent from the $G_0W_0$ quasiparticle energies.

In qs$GW$ the situation is similar to $G_0W_0$.  Equation~(\ref{eq:QPE2}) is also solved for the 
qs$GW$ quasiparticle energies. However, $v_{\rm xc}$ is replaced by $\tilde{\Sigma}$, the 
self-consistently determined, optimal, non-local, static potential that best represents 
the $G_0W_0$ self-energy. Since $\tilde{\Sigma}$ derives from $\Sigma^{G_0W_0}$ it also 
does not contain an interacting kinetic energy contribution and neither does $\epsilon^{\rm SP}_{n\sigma}$. 
The kinetic energy contribution is therefore also absent from the quasiparticle energies in the qs$GW$ framework.

We therefore conclude that although sc$GW$ and qs$GW$ at first glance appear to be similar 
$GW$ self-consistency schemes, they differ quite considerably in their treatment of the 
kinetic energy. We attribute the observed, average deviation of $\sim$0.45~eV between 
these two schemes to the difference in the kinetic energy treatment, the difference in the electron density and the screening properties. 

\subsection{Partially self-consistent \textit{GW}}

We now turn to the partially self-consistent $GW_0$ scheme. Unlike sc$GW$ and qs$GW$, the ionization energies of this partially self-consistent scheme still exhibit a dependence on the starting point, owing to the non-self-consistent treatment of $W$.\cite{marom/2012} To account for this dependence, we based our sc$GW_0$ calculations on two different starting points: PBE and HF. Our calculations for the $GW$100 set indicate that sc$GW_0$@PBE underestimate the ionization energies by 0.34~eV [Fig.~\ref{fig:error}~(c)], whereas sc$GW_0$@HF overestimates them by 0.06~eV [Fig.~\ref{fig:error}~(d)]. 
This trend reflects the over- and under-screening of the screened Coulomb interaction induced by the evaluation of $W$ with PBE or HF orbitals, respectively. In practice, owing to the band-gap problem of Kohn-Sham DFT\cite{Perdew1983} PBE calculations typically underestimate the HOMO-LUMO gap by as much as 50\% as compared to quantum-chemical calculations or reference experimental values. The small HOMO-LUMO gap, in turn, leads to an overestimation of the polarizability [Eq.~(\ref{eq:chi})] and, correspondingly, of the correlation part of the $G_0W_0$ self-energy, as alluded to in the previous Section. Conversely, HOMO-LUMO gaps are typically overestimated in Hartree-Fock owing to the lack of electronic correlation which leads, following similar arguments, to an underscreening of the polarizability and a corresponding overestimation of the quasiparticle energies. 

The sc$GW_0$@HF and sc$GW_0$@PBE ionization energies differ from each other  by 0.4~eV on average, with a maximum deviation of 1~eV (e.g., for F$_2$Mg). sc$GW_0$@HF exhibits the lowest MAE (0.2~eV) relative to CCSD(T) among the $GW$ methods considered in this work [Fig.~\ref{fig:abserror}~(d)].  It gives larger ionization energies than sc$GW$ on average. Since also the partial self-consistency scheme incorporates the interacting kinetic energy through the self-consistent Green's function, we attribute the larger ionization energies in sc$GW_0$@HF to a more pronounced underscreening due to the fact that the HF HOMO-LUMO gap that determines the screening strength of $W$@HF is larger than that of sc$GW$. Conversely, the underscreening in sc$GW_0$@PBE indicates that PBE-based screening ($W$@PBE) is not as suitable for the $GW$100 set as Hartree-Fock based screening ($W$@HF), although for larger molecules or solids, the situation may differ.  

\subsection{The perturbative \textit{G}$_0$\textit{W}$_0$ scheme}

For comparison, we report in Figs.~\ref{fig:error} and \ref{fig:abserror} the ME and MAE of $G_0W_0$@HF and $G_0W_0$@PBE. Other $G_0$ starting points will be discussed in connection to the DSLE-scheme in Section~\ref{sec:DSLE}.
$G_0W_0$@PBE underestimates the ionization energies by $~0.7$~eV [Fig.~\ref{fig:error} (e)], whereas $G_0W_0$@HF overestimates by $~0.3$~eV [Fig.~\ref{fig:error} (f)]. 
$G_0W_0$ calculations exhibit a more pronounced dependence on the starting point as compared to sc$GW_0$, since neither $G$ and $W$ are treated self-consistently.
The average discrepancy between $G_0W_0$@PBE and $G_0W_0$@HF ionization energies is approximately 1~eV, and can be as large as 2~eV. For $G_0W_0$, in particular, HF provides a better starting point as it leads to a mean absolute error a factor of 2 smaller as compared to PBE [Fig.~\ref{fig:abserror} (e)-(f)]. A similar observation was also made for the ionization energies and electron affinities of organic acceptor molecules. \cite{Knight/etal:2016}

As alluded to in Section~\ref{sec:scGWvsqsGW}, $G_0W_0$@HF gives results that are comparable to qs$GW$. However, sc$GW$ differs appreciably. Looking at the progression from $G_0W_0$@HF to sc$GW_0$@HF to sc$GW$ we can now understand the reduction of the ionization energies in terms of changes to the electronic screening, the electron density and the kinetic energy. Going from $G_0W_0$@HF to sc$GW_0$@HF incurs a density change as illustrated in Fig.~\ref{fig:dens} and a change from the non-interacting to the interacting kinetic energy (albeit without possible kinetic energy changes due to changes in $W$). Both effects together reduce the ionization energies on average. Going from sc$GW_0$@HF to sc$GW$ does not change the density appreciably anymore according to Fig.~\ref{fig:dens}. The additional reduction of the ionization energies in sc$GW$ therefore results from a reduction of the underscreening in $W$ in going from $W$@HF to the self-consistent $W$ and a concomitant change in the kinetic energy.

\subsection{Trends across the \textit{GW}100 set}

For all molecules of the $GW$100 set, the deviation from the CCSD(T) ionization energies is illustrated in Fig.~\ref{fig:deviation}. The horizontal shaded area marks points differing by less than 0.3~eV from CCSD(T).
As a guide through the chemical composition of the different compounds, we divided the $GW$100 set into ten subgroups: atoms, dimers, hydrocarbons, hydrides, halogenides, nitrides, oxides, aromatic molecules, nucleobases, and transition metals compounds. These categories are intended as an approximate indication of the chemical compositions of the $GW$100 subsets. Different categories are color-coded and separated by vertical dotted lines. 

Fig.~\ref{fig:deviation}~(b) shows that sc$GW$ provides accurate ionization energies for molecules of the hydride, halogenide, and oxide groups. The MAE reduces to $\sim$~0.15~eV if we consider only molecules of the oxide group. A common element of these compounds is the presence of highly electronegative atoms (O, F, Cl) and, correspondingly, the formation of covalent bonds with a strong ionic character. The largest discrepancies among the sc$GW$ ionization energies are observed for systems characterized predominantly by delocalized $\pi$-type orbitals such as, e.g., compounds of the hydrocarbon and nucleobase groups. At variance with sc$GW$, sc$GW_0$@HF [Fig.~\ref{fig:deviation}~(a)]  and $G_0W_0$@HF [Fig.~\ref{fig:deviation}~(b)] exhibit the largest deviation from CCSD(T) for ionic compounds (hydrides, halogenides, and oxides), whereas the discrepancy is small for $\pi$-orbital compounds. Figure~\ref{fig:deviation} further reveals that sc$GW_0$@PBE deviates rather homogeneously from the CCSD(T) reference data.

\section{DSLE-min \textit{GW} ionization energies}\label{sec:DSLE}

\begin{figure}[t]
  \includegraphics[width=0.68\textwidth]{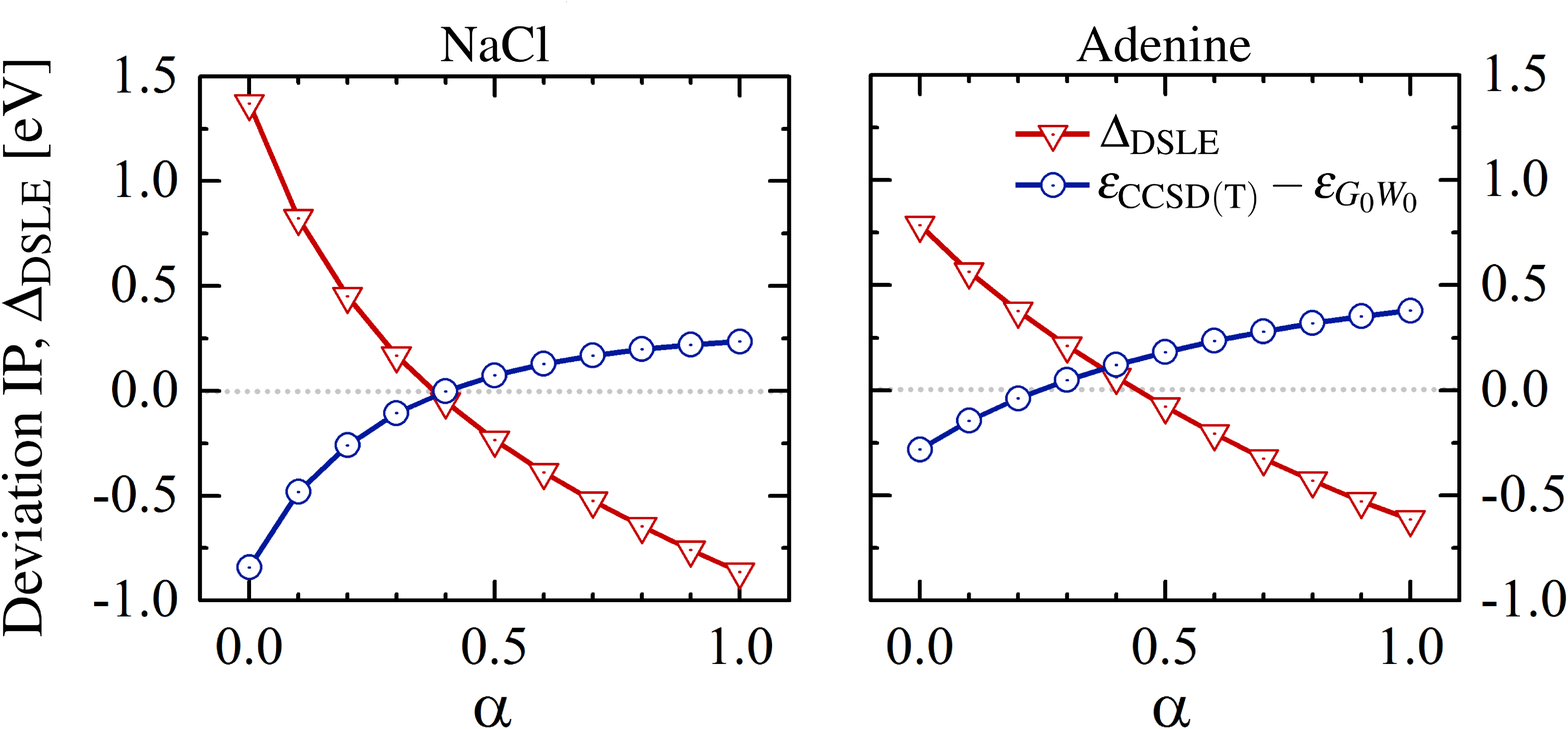}
  \caption{ 
  Correlation between DSLE and accuracy of the ionization energy for the NaCl  (left) and the adenine (right) molecules. The deviation  of the$G_0W_0$@PBEh($\alpha$) HOMO energies from the reference CCSD(T) ionization energies, $\epsilon_{\rm CCSD(T)} - \epsilon_{G_0W_0}$, is displayed in blue for different amounts of Hartree-Fock exchange $\alpha$ used in the PBE hybrid starting point. The $\Delta_{\rm DSLE}$ values are depicted in red as a function of $\alpha$. { We use Tier~4$^+$ basis sets for our DSLE-min $G_0W_0$ calculations.}}
  \label{fig:DSLE_molec}
  \end{figure}

We now turn to the discussion of the accuracy of basis-set converged (T4+) DSLE-min $GW$ calculations. 
In Fig.~\ref{fig:DSLE_molec}, we report the DSLE for two representative 
molecules of the $GW$100 test set, sodium chloride (left) and the adenine nucleobase (right).
In practice, the DSLE is estimated by evaluation of Eq.~(\ref{eq:dsle}) with the QP energies from $G_0W_0$@PBEh($\alpha$) 
for $\alpha$ between 0 (pure PBE exchange) and 1 (pure Hartree-Fock exchange). In addition, we show the
 deviation of the quasiparticle energy for the HOMO from the CCSD(T) reference 
($\epsilon_{\rm CCSD(T)}^{\rm HOMO} - \epsilon_{\rm QP}^{\rm HOMO}$). 
Both molecules exhibit a clear correlation between the DSLE and the accuracy of the ionization energy. 
Figure~\ref{fig:DSLE_molec} reveals that $\alpha$ values smaller than 0.4 typically result in a positive $\Delta_{\rm DSLE}$ and a corresponding underestimation of the ionization energy, whereas the opposite trend is observed for larger $\alpha$ values. At $\alpha\approx 0.4$ for NaCl and $\alpha\approx 0.45$ for adenine
we find $\Delta_{\rm DSLE}=0$. 
For NaCl, the DSLE-minimized starting point 
yields an ionization energy that coincides with the CCSD(T) result, whereas for adenine it  is slightly overestimated. More generally, we find  for all the systems in the $GW$100 set that the deviation from CCSD(T) is strongly reduced when the DSLE is minimized.

\begin{figure}[t]
  \includegraphics[width=0.48\textwidth]{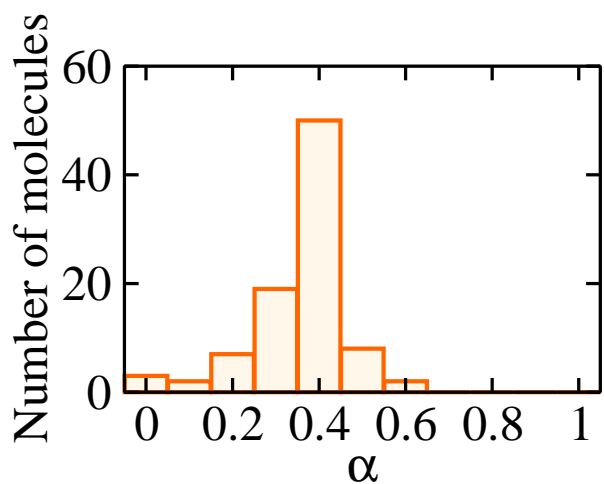}
  \caption{
Distribution of the optimal $\alpha$ values obtained from the DSLE-min $G_0W_0$ approach
for the GW100 test set with the Tier~4$^+$ basis set. }
  \label{fig:alpha}
  \end{figure}

\begin{figure}[t]
  \includegraphics[width=0.68\textwidth]{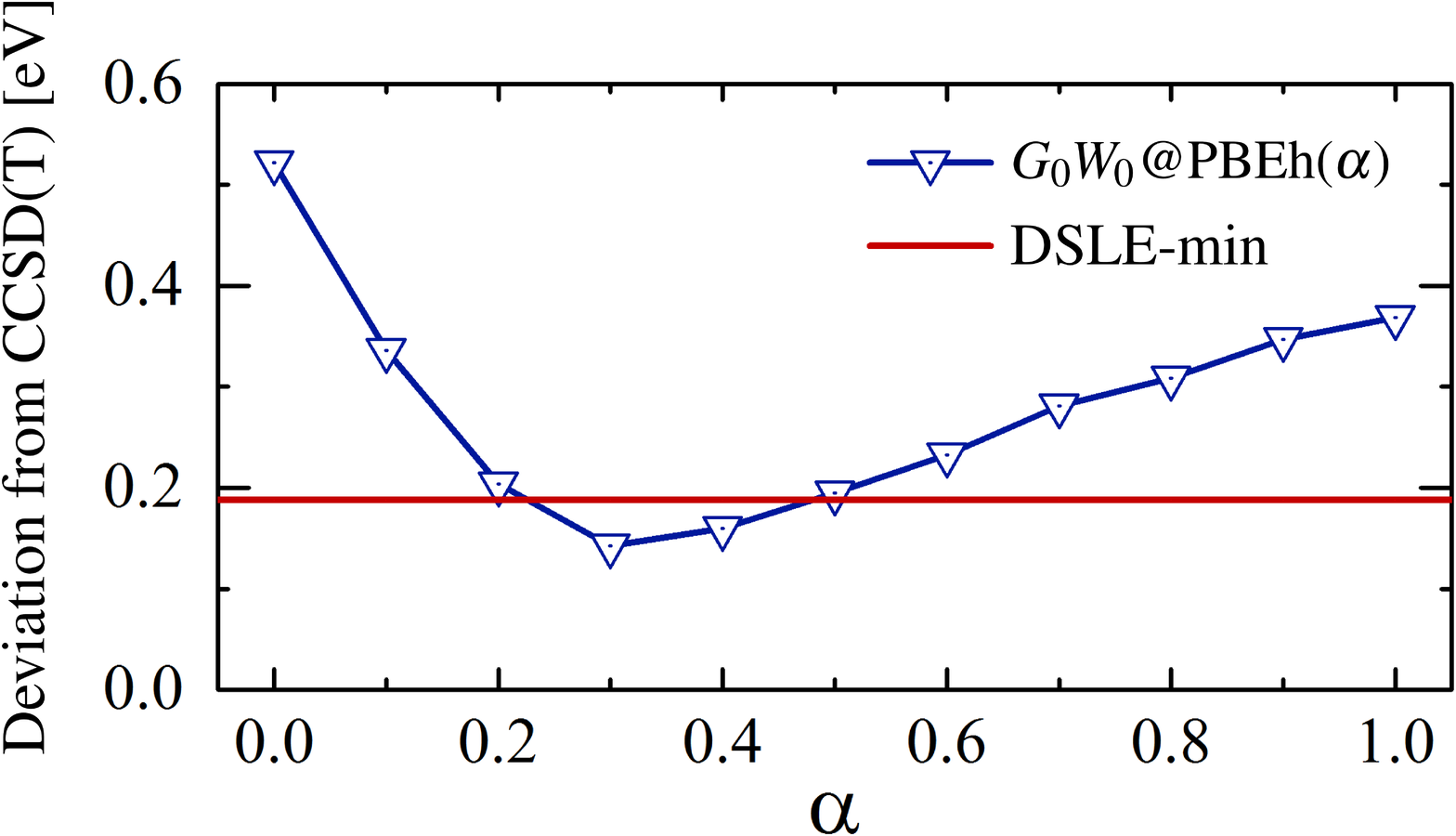}
  \caption{
Mean absolute error of the deviation from CCSD(T) for the $G_0W_0$@PBE($\alpha$)
ionization energies of the $GW$100 benchmark set as a function of $\alpha$. 
The MAE of DSLE-min $GW$ is reported as a red solid line. { We use Tier~4$^+$ basis sets for our DSLE-min $G_0W_0$ calculations.}}
  \label{fig:DSLE_gw100}
  \end{figure}

More generally, we find that also for other systems of the
$GW$100 set the deviation from CCSD(T) is strongly reduced whenever the DSLE is minimized.
In Fig.~\ref{fig:alpha}, we illustrate the distribution of optimal $\alpha$ values 
across the systems of the GW100 testset computed with the Tier~4$^+$  basis set. 
The optimal $\alpha$ determined from the 
DSLE-min $G_0W_0$ approach is almost unaffected by finite 
basis set errors owing to cancellation effects in Eq.~(\ref{eq:dsle}).
Only three molecules of the GW100 testset minimize the DSLE 
already for $\alpha=0$ (that is, for pure PBE exchange): LiH, Li$_2$, and Na$_2$.
The average over all $\alpha$ values amounts to 0.35.
This substantiates the results of previous starting-point benchmarks \cite{Bruneval2012,Korzdorfer2012c,Ko2014,Govoni2015}, 
which find a similar fraction of Fock exchange to provide the most accurate vertical ionization energies.
Figure~\ref{fig:DSLE_gw100} 
explicitly shows the MAE for the ionization energies of the $GW$100 set
obtained from $G_0W_0$@PBE($\alpha$) as a function of $\alpha$ and, marked by a horizontal red line,
the MAE of DSLE-min $GW$.

Finally, in Fig.~\ref{fig:DSLE_gw100} we report the MAE for the ionization energies of the $GW$100 set 
obtained from $G_0W_0$@PBE($\alpha$) as a function of $\alpha$ and, marked by a horizontal red line, 
the MAE of DSLE-min $GW$.
Figure~\ref{fig:DSLE_gw100} reveals that, 
among all possible choices of PBEh($\alpha$) starting points, the 
DSLE-minimization procedure yields a gratifying MAE and, thus, is a reliable choice for ionization energy predictions.

\section{Conclusions}\label{sec:conc}

In summary, we have studied the accuracy of 
state-of-the-art techniques based on many-body perturbation theory for 
the description of (charged) electronic excitations in molecules. 
For compounds of the $GW$100 benchmark set, we have computed the 
ionization energies as obtained from perturbative ($G_0W_0$) and 
self-consistent $GW$ approaches (sc$GW$, qs$GW$, and sc$GW_0$), as well from the recently developed DSLE-min $GW$ approach. 
Based on the comparison with CCSD(T) reference data, 
the results presented here quantify the overall accuracy 
of different flavors of $GW$ calculations for molecular 
compounds of diverse chemical composition. 
Overall, our sc$GW$ calculations suggest that the effect of vertex correction may 
become important for compounds characterized by chemical bonds with a pronounced 
ionic character (as, for instance, halogenides) or by nitrogen-lone pair orbital types, 
as these compounds exhibit the largest deviation from CCSD(T). Conversely, 
 sc$GW$ ionization energies lie typically within 0.3~eV from CCSD(T) for 
covalently bonded compounds.
The comparison between sc$GW$, sc$GW_0$, and qs$GW$ further reveals that 
different forms of self-consistency may influence the ionization energies  
and its agreement with the reference data considerably. We have identified underscreening, density changes and the treatment of the kinetic energy as reasons for the difference in the different self-consistent $GW$ schemes.
Finally, we have shown that the deviation from CCSD(T) may in part be attributed to the DSLE and, correspondingly, the DSLE-minimization procedure recently proposed by some of the authors emerges as a promising way to optimize the starting point of $G_0W_0$ calculations to improve the prediction of ionization energies.

\begin{acknowledgement}
We thank Xinguo Ren and Stephan K\"ummel for fruitful discussions. This work was supported by the Academy of Finland through its Centres of Excellence Programme under project numbers 251748 and 284621. M.D.\ acknowledges support by Deutsche Forschungsgemeinschaft 
Graduiertenkolleg 1640 and the Bavarian State Ministry of Science, Research, and the Arts for the Collaborative Research Network Soltech.
\end{acknowledgement}

%
%

\bibliography{references}

\begin{tocentry}
  \includegraphics[width=1\textwidth]{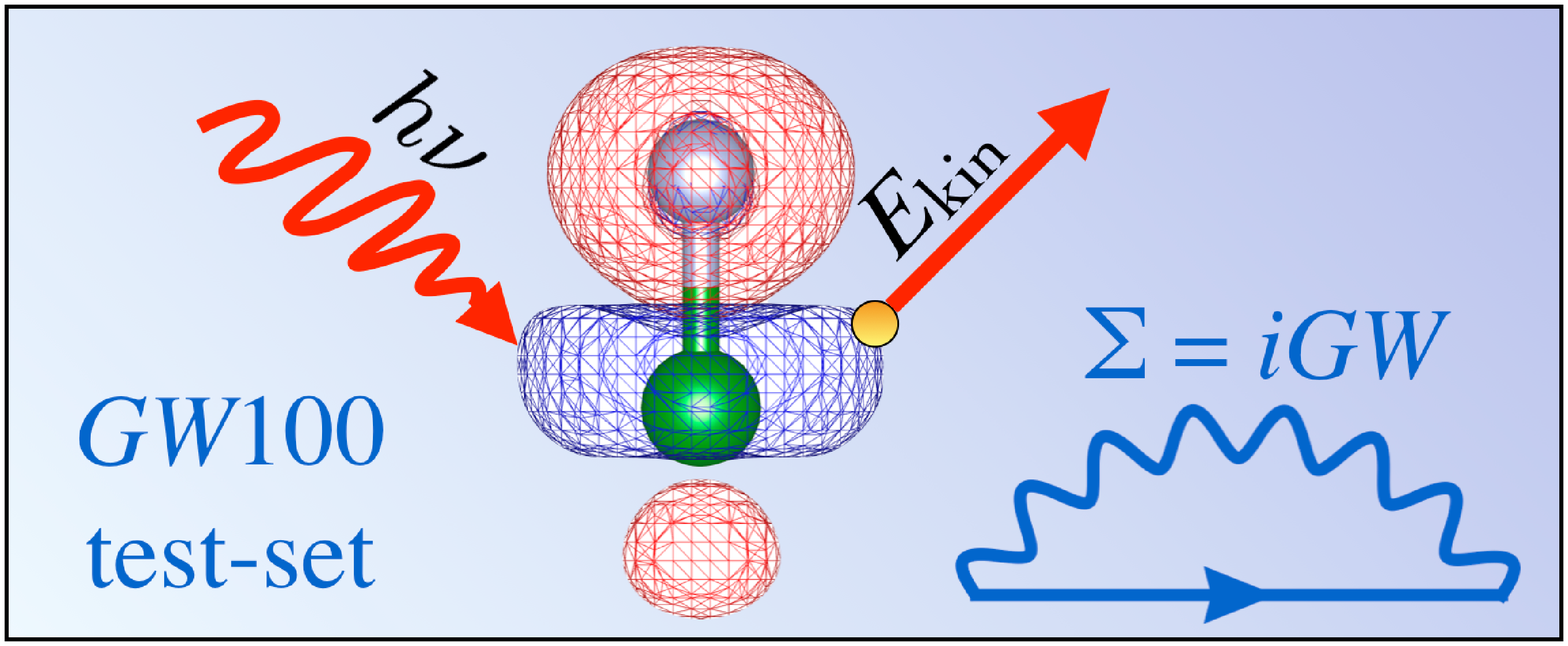}
\end{tocentry}

\end{document}